\documentclass{aip-cp}

\usepackage[numbers]{natbib}
\usepackage{rotating}
\usepackage{subcaption}
\usepackage{txfonts,graphicx}
\usepackage{textcomp}

\newcommand{\utilde}{\raise.17ex\hbox{$\scriptstyle\mathtt{\sim}$}}

\begin{document}

\title{Inauguration and First Light of the GCT-M Prototype for the Cherenkov Telescope Array}

\author[aff6]{J.J.~Watson\corref{cor1}}
\author[aff6]{A.~De Franco}
\eaddress{Andrea.DeFranco@physics.ox.ac.uk}
\author[aff1]{A.~Abchiche}
\author[aff2]{D.~Allan}
\author[aff3]{J.-P.~Amans}
\author[aff2]{T.P.~Armstrong}
\author[aff4]{A.~Balzer}
\author[aff4]{D.~Berge}
\author[aff3]{C.~Boisson}
\author[aff3]{J.-J.~Bousquet}
\author[aff2]{A.M.~Brown}
\author[aff4]{M.~Bryan}
\author[aff1]{G.~Buchholtz}
\author[aff2]{P.M.~Chadwick}
\author[aff5]{H.~Costantini}
\author[aff6]{G.~Cotter}
\author[aff7]{M.K.~Daniel}
\author[aff3]{F.~De Frondat}
\author[aff3]{J.-L.~Dournaux}
\author[aff3]{D.~Dumas}
\author[aff5]{J.-P.~Ernenwein}
\author[aff3]{G.~Fasola}
\author[aff8]{S.~Funk}
\author[aff1,aff3]{J.~Gironnet}
\author[aff2]{J.A.~Graham}
\author[aff7]{T.~Greenshaw}
\author[aff3]{O.~Hervet}
\author[aff9]{N.~Hidaka}
\author[aff10]{J.A.~Hinton}
\author[aff3]{J.-M.~Huet}
\author[aff3]{I.~Jegouzo}
\author[aff8]{T.~Jogler}
\author[aff8]{M.~Kraus}
\author[aff11]{J.S.~Lapington}
\author[aff3]{P.~Laporte}
\author[aff3]{J.~Lefaucheur}
\author[aff4]{S.~Markoff}
\author[aff3]{T.~Melse}
\author[aff8]{L.~Mohrmann}
\author[aff11]{P.~Molyneux}
\author[aff2]{S.J.~Nolan}
\author[aff9]{A.~Okumura}
\author[aff11]{J.P.~Osborne}
\author[aff10]{R.D.~Parsons}
\author[aff11]{S.~Rosen}
\author[aff11]{D.~Ross}
\author[aff12]{G.~Rowell}
\author[aff13]{C.B.~Rulten}
\author[aff9]{Y.~Sato}
\author[aff3]{F.~Say\`{e}de}
\author[aff2]{J.~Schmoll}
\author[aff10]{H.~Schoorlemmer}
\author[aff3]{M.~Servillat}
\author[aff3]{H.~Sol}
\author[aff12]{V.~Stamatescu}
\author[aff4]{M.~Stephan}
\author[aff14]{R.~Stuik}
\author[aff11]{J.~Sykes}
\author[aff9]{H.~Tajima}
\author[aff11]{J.~Thornhill}
\author[aff10]{L.~Tibaldo}
\author[aff5]{C.~Trichard}
\author[aff4]{J.~Vink}
\author[aff10]{R.~White}
\author[aff9]{N.~Yamane}
\author[aff3]{A.~Zech}
\author[aff8]{A.~Zink}
\author[aff10]{J.~Zorn}
\author{the CTA Consortium}
\eaddress[url]{http://www.cta-observatory.org}
\affil[aff6]{Department of Physics, University of Oxford, Keble Road, Oxford OX1 3RH, UK}
\affil[aff1]{CNRS, Division technique DT-INSU, 1 Place Aristide Briand, 92190 Meudon, France}
\affil[aff2]{Department of Physics and Centre for Advanced Instrumentation, Durham University, South Road, Durham DH1 3LE, UK}
\affil[aff3]{Observatoire de Paris, CNRS, PSL University, LUTH \& GEPI, Place J. Janssen, 92195, Meudon cedex, France}
\affil[aff4]{GRAPPA, University of Amsterdam, Science Park 904, 1098 XH Amsterdam, The Netherlands}
\affil[aff5]{Aix Marseille Universit\'{e}, CNRS/IN2P3, CPPM UMR 7346 ,  163 avenue de Luminy, case 902, 13288 Marseille, France}
\affil[aff7]{University of Liverpool, Oliver Lodge Laboratory, P.O. Box 147, Oxford Street, Liverpool L69 3BX, UK}
\affil[aff8]{Erlangen Center for Astroparticle Physics (ECAP), Erwin- Rommel-Str. 1, D 91058 Erlangen, Germany}
\affil[aff9]{Institute for Space-Earth Environmental Research, Nagoya University, Furo-cho, Chikusa-ku, Nagoya, Aichi 464-8601, Japan}
\affil[aff10]{Max-Planck-Institut f\"{u}r Kernphysik, P.O. Box 103980, D 69029 Heidelberg, Germany}
\affil[aff11]{Department of Physics and Astronomy, University of Leicester, University Road, Leicester, LE1 7RH, UK}
\affil[aff12]{School of Physical Sciences, University of Adelaide, Adelaide5005, Australia}
\affil[aff13]{Department of Physics and Astronomy, University of Minnesota, 116 Church Street, Minneapolis, MN, 55455, U.S.A.}
\affil[aff14]{Leiden Observatory, Leiden University, Postbus 9513, 2300 RA, Leiden, Netherlands}
\corresp[cor1]{Corresponding author: Jason.Watson@physics.ox.ac.uk}

\maketitle

\begin{abstract}
The Gamma-ray Cherenkov Telescope (GCT) is a candidate for the Small Size Telescopes (SSTs) of the Cherenkov Telescope Array (CTA). Its purpose is to extend the sensitivity of CTA to gamma-ray energies reaching 300 TeV. Its dual-mirror optical design and curved focal plane enables the use of a compact camera of 0.4 m diameter, while achieving a field of view of above 8 degrees. Through the use of the digitising TARGET ASICs, the Cherenkov flash is sampled once per nanosecond continuously and then digitised when triggering conditions are met within the analogue outputs of the photosensors. Entire waveforms (typically covering 96 ns) for all 2048 pixels are then stored for analysis, allowing for a broad spectrum of investigations to be performed on the data. Two prototypes of the GCT camera are under development, with differing photosensors: Multi-Anode Photomultipliers (MAPMs) and Silicon Photomultipliers (SiPMs). During November 2015, the GCT MAPM (GCT-M) prototype camera was integrated onto the GCT structure at the Observatoire de Paris-Meudon, where it observed the first Cherenkov light detected by a prototype instrument for CTA.
\end{abstract}

\section{INTRODUCTION}
The Cherenkov Telescope Array (CTA) will be the next generation of IACT (Imaging Atmospheric Cherenkov Telescope) array, improving on current Cherenkov instruments with a factor of 10 improvement in sensitivity. It will observe in the energy range 20 GeV - 300 TeV with a large field of view for surveys, improved angular and energy resolution, and will be the first IACT array to operate as an open observatory \citep{Acharya2013}.

CTA will consist of both a north and a south site,  with La Palma (Spain) and Cerro Amazons (Chile)  selected respectively for ongoing negotiations. Utilising around 100 telescopes, the south site will observe the full energy range split across three categories of telescopes: Large Size Telescope (LST), Medium Size Telescope (MST), and Small Size Telescope (SST). The purpose of the SSTs is to observe the highest gamma-ray energies of $E_\gamma>5$ TeV, where the showers create an abundance of Cherenkov photons, but are rarer. In order to observe a significant amount of these showers, a large quantity of SST telescopes are spread across an area greater than 5 km$^2$, however only a modest primary mirror diameter of \utilde4 m is necessary. Approximately two-thirds of the southern array telescopes will be SSTs, whereas there are currently no plans for SSTs in the northern site.

The Gamma-ray Cherenkov Telescope (GCT) is one of three proposed telescope designs for the SSTs. Designed and built by an Australian-Dutch-French-German-Japanese-UK-US consortium, the GCT uses a dual mirror Schwarzschild–
Couder (SC) optical design in order to increase FoV while reducing the required camera size. Two prototype cameras are under development for GCT: one using Multi-Anode Photomultiplier Tubes (MAPMTs) known as GCT-M, and another using Silicon Photomultiplier Tubes (SiPMTs) known as GCT-S. This paper will briefly describe the GCT-M telescope, and the November 2015 campaign where it was inaugurated and recorded its first Cherenkov light at the Observatoire de Paris-Meudon.

\begin{figure}[h]
	\captionsetup{type=figure}
	\label{fig:lab}
	\caption{The GCT-M camera testing in Leicester, UK.}
	\begin{subfigure}[b]{0.49\textwidth}
		\includegraphics[width=\textwidth]{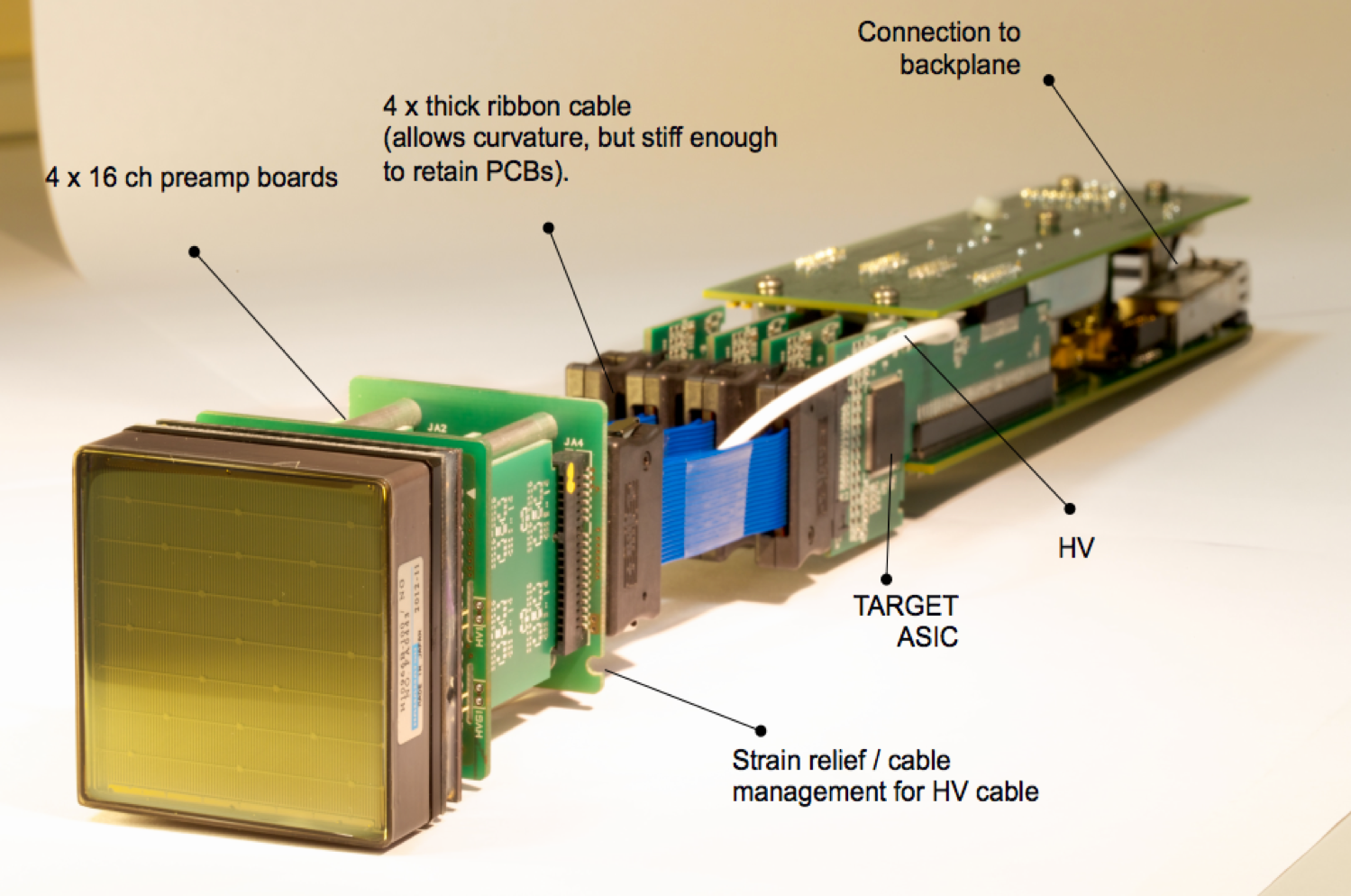}
		\caption{A GCT-M module, containing the MAPMT, pre-amplifier and readout electronics.}
	\end{subfigure}
	~
	\begin{subfigure}[b]{0.49\textwidth}
		\includegraphics[width=\textwidth]{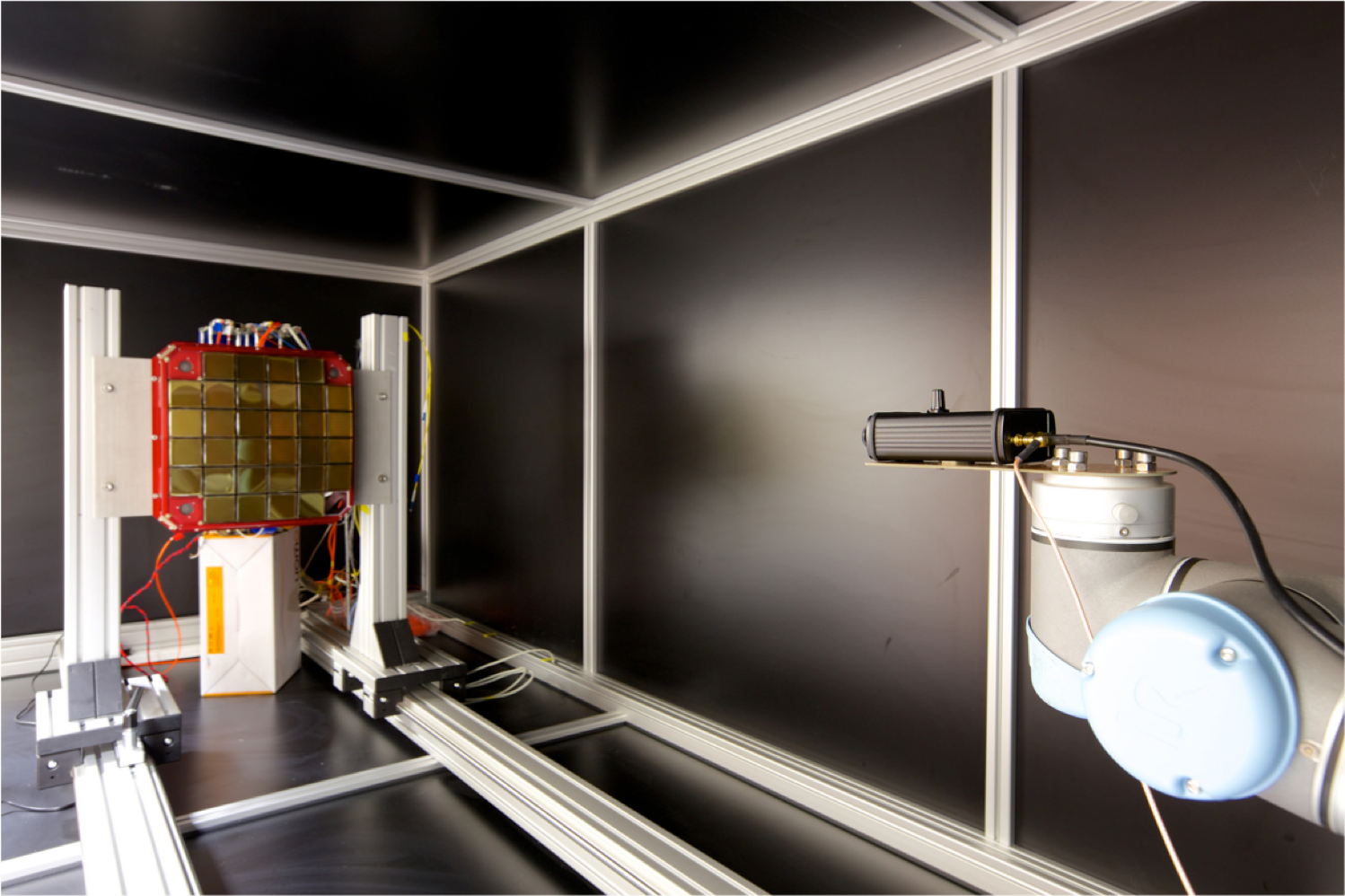}
		\caption{The GCT-M test bench.}
	\end{subfigure}
\end{figure}


\section{THE GCT-M TELESCOPE}
\begin{figure}[h]
	\captionsetup{type=figure}
	\caption{\label{fig:telescope} The GCT-M telescope with the camera mounted at the Observatoire de Paris-Meudon.}
	\begin{subfigure}[b]{0.49\textwidth}
		\includegraphics[width=\textwidth]{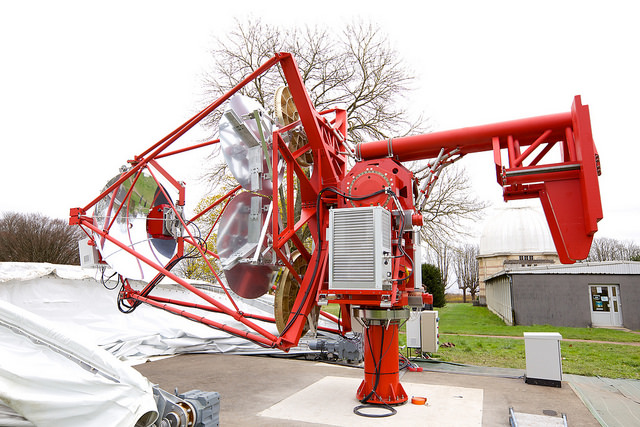}
	\end{subfigure}
	~
	\begin{subfigure}[b]{0.49\textwidth}
		\includegraphics[width=\textwidth]{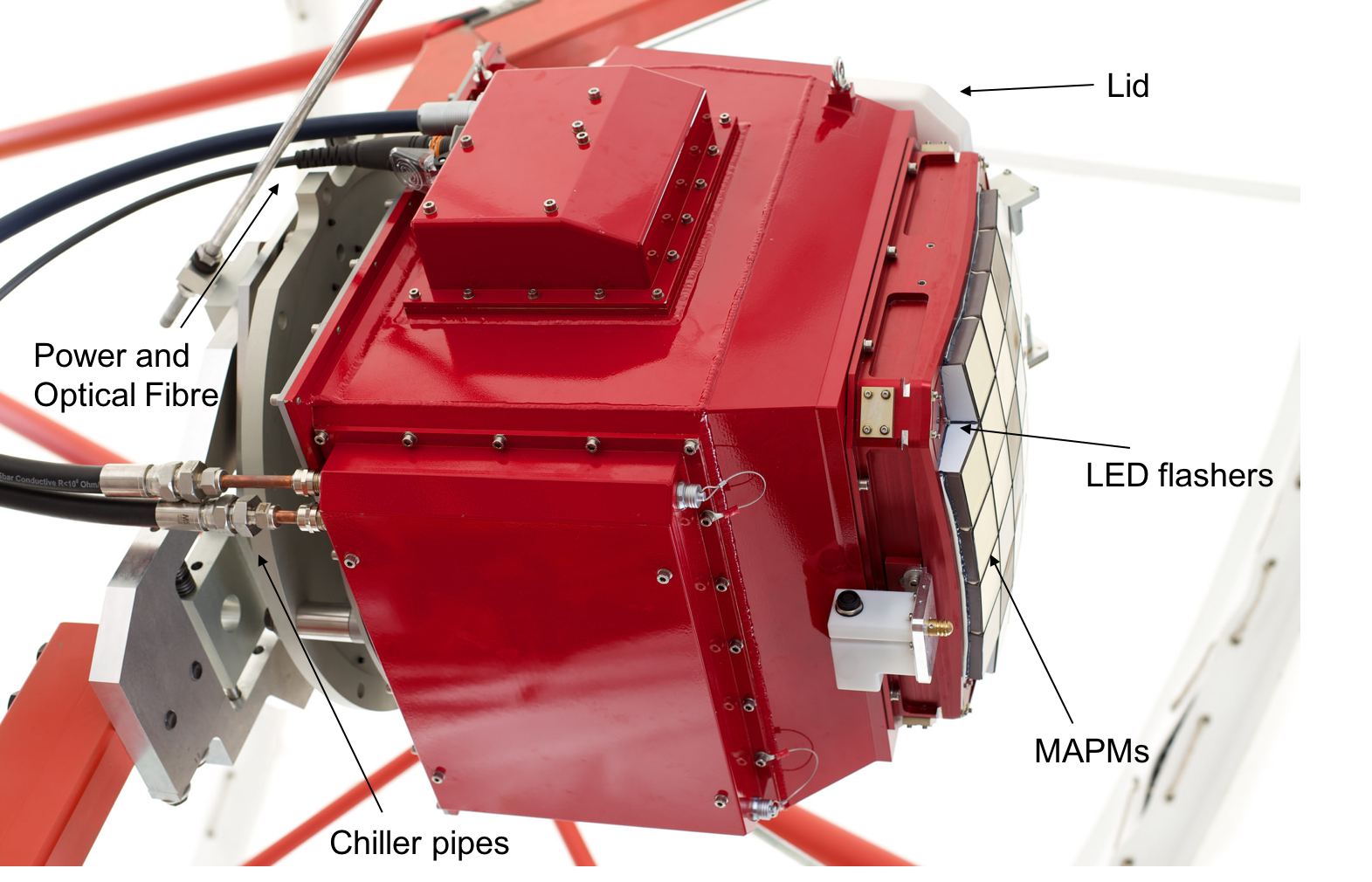}
	\end{subfigure}
\end{figure}

The GCT-M prototype can be reduced to the following components (for more descriptive details on the telescope please refer to the proceedings by L. Tibaldo et al. in this issue):

\begin{itemize}
\item \textbf{Telescope Structure:}
A Dual-mirror Schwarzchild-Couder design, enabling a large FoV and good angular resolution, while reducing the focal length thereby allowing the construction of a more compact and lightweight camera and telescope structure \cite{Dournaux2016}. See the proceedings by H. Constantini et al. in this issue for further information on the telescope structure.	
\item \textbf{Multi-Anode Photomultiplier Tubes:}
Photosensors consisting of tightly packed photomultipliers. Each module contains 64 pixels. 32 modules are installed per camera.
\item \textbf{Pre-Amplifier:}
Amplifies and shapes the photomultiplier signal.
\item \textbf{TARGET Module:}
Sample per nanosecond digitisation device. Also forms front end trigger. Refer to the proceedings by D. Jankowsky et al. in this issue for more details about the TARGET module.
\item \textbf{Backplane:}
Handles the front end trigger, and routes raw data to two data acquisition boards (DACQ). Also distributes power and controls housekeeping. 
\item \textbf{DACQ Boards:}
Serialise data to an external PC with two 1Gbit/s optical fibre link per board. Also handles control of the backplane and peripherals.
\item \textbf{LED Flashers:}
Provide calibration through illumination of the focal plane over a range of intensities. Achieved via reflection from the secondary mirror. Installed in each corner of the camera face. See \citet{Brown2015} for more information.
\item \textbf{Cooling:}
The camera is liquid cooled via a chiller unit mounted on the telescope. Internal fans circulate air via a set of mechanical baffles.
\item \textbf{Optical Fibre:}
Camera readout and communications take place via optical fibre. 
\item \textbf{Power:}
The camera is powered by a single 12 V DC supply, mounted at the rear of the secondary mirror. The power consumption is less than 450 W.
\item \textbf{Lid:}
An external lid system provides protection from the elements. 
\item \textbf{Peripherals Board:}
Controls the lid, fans, LED flashers, and a set of internal sensors (temperature and humidity).
\end{itemize}
Figure \ref{fig:lab}a and Figure \ref{fig:telescope} display annotations of the camera's components.

\section{GCT-M COMMISSIONING AND INAUGURATION}
\begin{figure}[h]
	\centering
	\includegraphics[width=\linewidth]{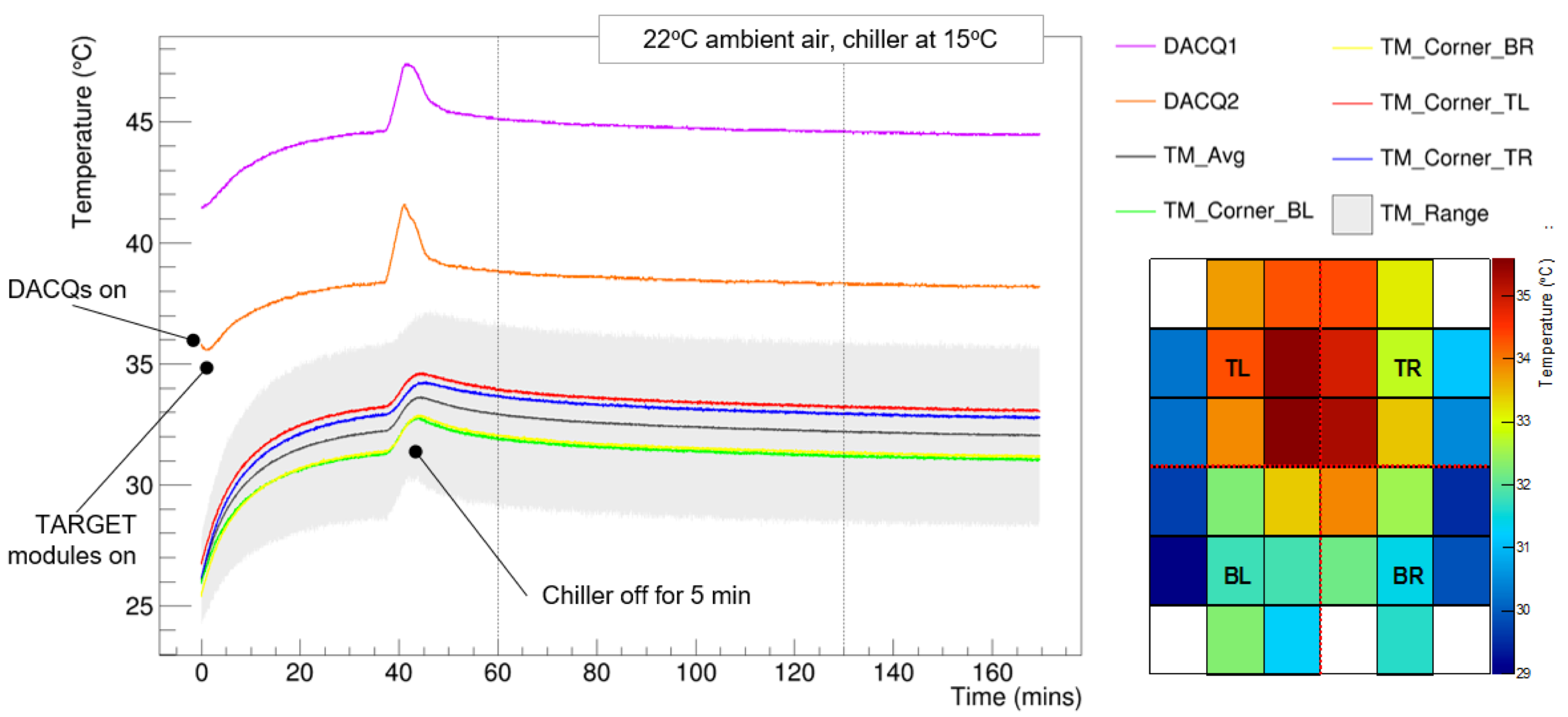}
	\caption{\label{fig:temp} Visualisation of the temperature monitoring over a 170 minute period, during which the effectiveness of the chiller was tested by switching it off for 5 minutes. In the top right is the legend for the graph lines, where ``TM" stands for TARGET module. Shown in the bottom right is the time-averaged temperature distribution across the camera face from 60 to 130 minutes. \cite{Brown2016}}
\end{figure}

Before installing the GCT-M camera on the telescope structure in Meudon, it was first commissioned in Leicester, UK. Using a light-tight container and a laser mounted upon a robot-arm (Figure \ref{fig:lab}b), a variety of tests were performed. These tests included arrival time calibration from pico-second laser flashes, single-photoelectron measurements for each pixel, gain-matching of all modules, and the determination of transfer-functions for all cells of the four ASICs located on each of the 32 TARGET modules \cite{Brown2016}. Additionally, the safety of the camera was evaluated through procedures such as temperature monitoring \& control, ensuring the camera was installation ready (Figure \ref{fig:temp}).

In November 2015 the camera was placed into a shock-absorbing shipping case and sent from Leicester to Paris. To ensure no damage occurred during transportation, tests were repeated on arrival before the installation onto the telescope structure. The mounting arm for the camera has the ability to be lowered, enabling the installation or removal of the camera to be accomplished in 5 steps, requiring 15 minutes and 2-3 people. This allows for swift on-site camera maintenance. Once installed, camera operation on the telescope was conducted with extreme precaution; to ensure the MAPMTs were not exposed to excessive light levels the high voltage was initially only activated with the lid and shelter closed, and then gradually exposed to greater light levels when the sun had set. Once again, camera safety checks were performed, and first on-telescope data acquisition was tested using the reflection of the calibration LEDs on the secondary mirror. A photo of the camera mounted on the telescope is shown in Figure \ref{fig:telescope}.

\begin{figure}[h]
	\centering
	\includegraphics[width=0.9\linewidth]{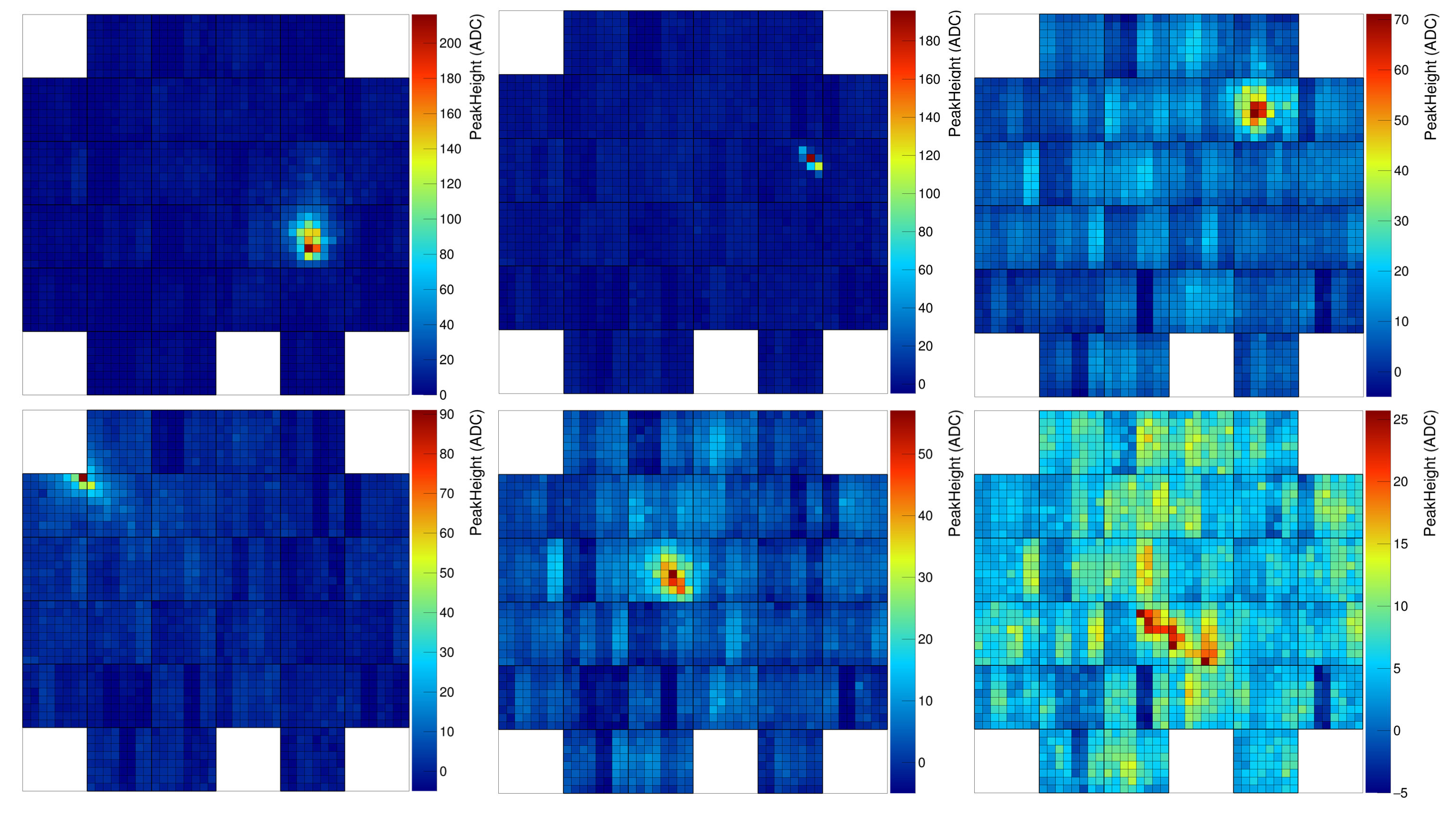}
	\caption{\label{fig:events} A selection of the first Cherenkov shower events detected by GCT-M. The intensity in each pixel is the maximum value of that waveform's pixel after pedestal subtraction. The missing module is a result of a faulty connector on the backplane that has since been replaced. The last event (bottom-right) occurred when the cloud covering the moon reduced, increasing the NSB (Night-Sky Background).}
\end{figure}

\begin{figure}[hb]
	\centering
	\includegraphics[width=0.9\linewidth]{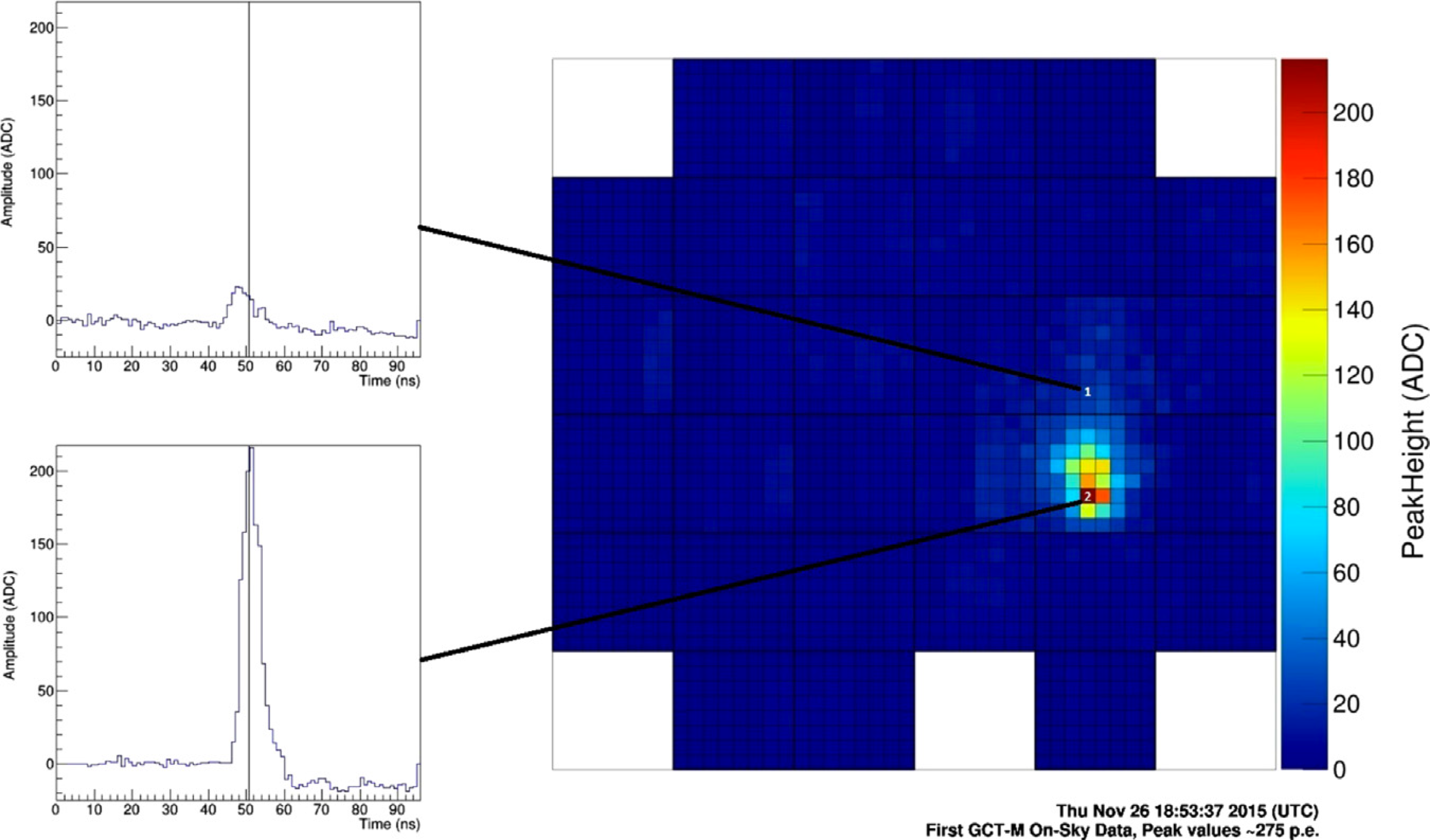}
	\caption{\label{fig:movie} A frame of an animation showing the progression of the pedestal-subtracted event. The frame coincides with the time of maximum pixel intensity (vertical black line on the waveform). The full animation can be found in the Gamma2016 presentation by L. Tibaldo.}
\end{figure}

On the 26th of November 2015, the telescope was ready to attempt on-sky observations. Unfortunately the weather conditions had been poor, and the moon was almost full. However, an opportunity was found where the skies were clear, apart from some cloud covering the moon. Nevertheless the NSB was at least 500~MHz (p.e./sec/pixel), about 50 times brighter than is expected at the CTA southern site in Chile \cite{Dournaux2016}. In order to operate at this NSB without any risk to the camera, the high voltage was only operated at 750~V, lower than the typical laboratory values of 950~V and 1100~V, consequently the gain calibration for this observation was not known. Additionally only 2 of the 6 primary mirrors had reflective surfaces, the other 4 were dummies to provide the correct surface area and weight. None of the mirrors were aligned. Despite these difficulties, while using a high trigger threshold around 30 Cherenkov showers were detected within \utilde10 minutes. A selection of these Cherenkov events can be seen in Figure \ref{fig:events}. The profile of the events closely resembles what is predicted from simulations, and are likely to be hadron-initiated Cherenkov showers (due to the absence of telescope pointing). Figure \ref{fig:movie} shows a frame of an animation of a Cherenkov event, with an approximated maximum intensity of 275 photoelectrons. Recording the whole waveform in each pixel is a powerful tool for further analysis of the showers \cite{Brown2016}; the waveforms demonstrate the expected few nanosecond duration of the event and occur in a coincident time-window in the relevant pixels. Additionally shown in the animation from Figure \ref{fig:movie}, the movement of the intensity across the camera is consistent with the signature characteristic of an air shower signal \cite{Dournaux2016}.

\clearpage

\section{SUMMARY}
The GCT-M camera was successfully commissioned and installed onto the telescope structure at the Observatoire de Paris-Meudon in late November. Due to some fortunate cloud cover in front of the moon, the telescope could be pointed safely at clear skies and begin data acquisition. During a 10 minute period around 30 Cherenkov showers were detected, likely with hadronic origin. These are the first Cherenkov showers to be detected by a CTA prototype. After some further lab-commissioning at the Max-Planck-Institut f\"{u}r Kernphysik in Heidelberg, the camera will return to Meudon for a second observation campaign in late 2016.

\section{ACKNOWLEDGEMENTS}
We acknowledge support from the EU 7th Framework Programme for research, technological development and demonstration under grant agreement n. 317446. We also gratefully acknowledge support from the agencies and organisations listed under Funding Agencies at this website: http://www.cta-observatory.org/.


\bibliographystyle{aipnum-cp}%
\bibliography{papers}%

\end{document}